\documentclass[11pt,letterpaper]{article}
\usepackage[top=0.85in,left=1.55in,footskip=0.75in,marginparwidth=2in]{geometry}

\usepackage[utf8]{inputenc}

\usepackage[sort,compress]{cite}

\usepackage{amsmath}

\usepackage[table]{xcolor}

\usepackage{nameref,hyperref}
\hypersetup{
    colorlinks=true,
    linkcolor=black,
    citecolor=black,
    filecolor=black,
    urlcolor=black,
}
\usepackage[right]{lineno}

\usepackage{microtype}
\DisableLigatures[f]{encoding = *, family = * }

\usepackage{changepage}

\usepackage[aboveskip=1pt,labelfont=bf,labelsep=period,singlelinecheck=off]{caption}

\makeatletter
\renewcommand{\@biblabel}[1]{\quad#1.}
\makeatother

\usepackage{lastpage,fancyhdr,graphicx}
\usepackage{epstopdf}
\fancyheadoffset[L]{2.25in}
\fancyfootoffset[L]{2.25in}

\usepackage{color}

\definecolor{Gray}{gray}{.25}

\usepackage{graphicx}

\usepackage{sidecap}

\usepackage{wrapfig}
\usepackage[pscoord]{eso-pic}
\usepackage[fulladjust]{marginnote}
\reversemarginpar

\begin{document}
\vspace*{0.35in}

\begin{flushleft}
{\Large
\textbf\newline{scX: A user-friendly tool for scRNA-seq exploration.}
}
\newline
\\
Tomás Vega Waichman\,$^{\text{1}}$, M. Luz Vercesi\,$^{\text{1}}$, Ariel A. Berardino\,$^{\text{1,2}}$, Maximiliano S. Beckel\,$^{\text{1,2}}$, Damiana Giacomini\,$^{\text{2,3}}$, Natalí B. Rasetto\,$^{\text{2,3}}$, Magalí Herrero\,$^{\text{2,3}}$, Daniela J. Di Bella\,$^{\text{4}}$, Paola Arlotta\,$^{\text{4}}$, Alejandro F. Schinder\,$^{\text{2,3}}$ and Ariel Chernomoretz\,$^{\text{1,5}*}$
\\

\bigskip
\begin{footnotesize}
$^1$ Integrative Systems Biology Lab, Leloir Institute, Buenos Aires, C1405 BWE, Argentina.

$^2$ Instituto de Investigaciones Bioqu\'{\i}micas de Buenos Aires, Consejo Nacional de Investigaciones Cient\'{\i}ficas y T\'ecnicas (CONICET), Buenos Aires, C1425 FQB, Argentina.

$^3$ Laboratory of Neuronal Plasticity, Leloir Institute, Buenos Aires, C1405 BWE, Argentina.

$^4$ Dept. of Stem Cells and Regenerative Biology, Harvard University \& Stanley Center for Psychiatric Research, Broad Institute of MIT and Harvard, Cambridge, MA, USA.

$^5$ Departamento de F\'{\i}sica, Facultad de Ciencias Exactas y Naturales, Universidad de Buenos Aires, Instituto de F\'{\i}sica de Buenos Aires, Consejo Nacional de Investigaciones Cient\'{\i}ficas y T\'ecnicas (CONICET), Buenos Aires, Argentina

\end{footnotesize}
\bigskip
\end{flushleft}

\abstract{Single-cell RNA sequencing (scRNA-seq) has transformed our ability to explore biological systems. Nevertheless, proficient expertise is essential for handling and interpreting the data. In this paper, we present scX, an R package built on the Shiny framework that streamlines the analysis, exploration, and visualization of single-cell experiments. With an interactive graphic interface, implemented as a web application, scX provides easy access to key scRNAseq analyses, including marker identification, gene expression profiling, and differential gene expression analysis. Additionally, scX seamlessly integrates with commonly used single-cell Seurat and SingleCellExperiment R objects, resulting in efficient processing and visualization of varied datasets. Overall, scX serves as a valuable and user-friendly tool for effortless exploration and sharing of single-cell data, simplifying some of the complexities inherent in scRNAseq analysis.}



\section{Introduction}
After nearly fifteen years of continuous development, single-cell transcriptomics continues to have a profound impact on the biomedical research field~\cite{Regev2017, Nomura2021, Suva2019, Rood2022, Sikkema2023}. 
Over the years, various data-processing pipelines have been proposed~\cite{Wolf2018, Butler2018, Luecken2019, Amezquita2020, Andrews2021}
, as well as visualization tools that aimed to ease the analysis of this type of high-throughput assays~\cite{Cakir2020}. 

The existing software ecosystem is extensive, often exhibiting overlap in approaches and solutions. Table ST1 (included as supplementary material) presents a comprehensive comparison of several commonly used tools for scRNAseq data exploration. This table provides insight into the extent to which each solution covers various aspects of the analysis. iSEE~\cite{Rue-Albrecht2018}, cellxgene~\cite{CZI_Single-Cell_Biology_Program2023},and ShinyCell~\cite{Ouyang2021} are noteworthy tools that primarily concentrate on data visualization, offering a diverse range of plots and graphical data representations. 
ShIVA~\cite{Aussel2023} and CellSnake~\cite{Umu2023} on the other hand are solutions more focused on data processing aspects of single cell analysis. There are also more comprehensive tools, such as ASAP~\cite{Fabrice2020}, SEQUIN~\cite{Weber2023}, and SCHNAPPS~\cite{Jagla2021} 
that can handle multiple embeddings and visualizations, and grant interactive single-cell analysis features such as clustering, marker identification, and differential expression analysis. 
Each one of them addresses differently the trade-off between the extend and complexity of the offered computational calculations and design criteria in terms of usability and ease of interaction with the application. 
This election has profound impacts on the user side. Specifying numerous parameters may pose challenges for users who are typically more interested in extracting relevant biology from their data and may lack the expertise or criteria to define the required values for each presented option. In such cases, while comprehensiveness is desirable, it may hinder or compromise the tool's ease of use.

Here we present scX, an R package that deploys a user-friendly Shiny-based application developed for researchers to explore single-cell datasets. 
From its inception, scX was designed as a tool to efficiently bridge the computational side of the problem (e.g. data preparation, normalization, markers identification, differential expression, etc) with various tools enabling the rapid implementation of biological analyses derived from these results (interactive 3D visualizations of low-dimensional embeddings, on-the-fly markers identification, exploratory data analysis capabilities, availability of a wide array of plot types, etc).
scX becomes particularly well suited for two scenarios. For one hand, for bioinformatics laboratories looking to share scRNAseq experiment data, processed with arbitrary sophistication, with biology colleagues aiming to explore them efficiently in a user-friendly designed platform. At the same time, our package offers, if required, the ability to carry out a significant portion of typical scRNA-seq data analysis (normalization, dimensionality reduction, clustering, marker identification, and calculation of differential expression) in a non-interactive preprocessing step that can complement or integrate with any existing analyses. This ensures that, even users new to computational aspects of this field have the opportunity to benefit from the streamlined analysis of their data through the tools provided by the interactive scX interface.

\begin{figure*}[t]
    \begin{center}
       \includegraphics[width=6.2in]{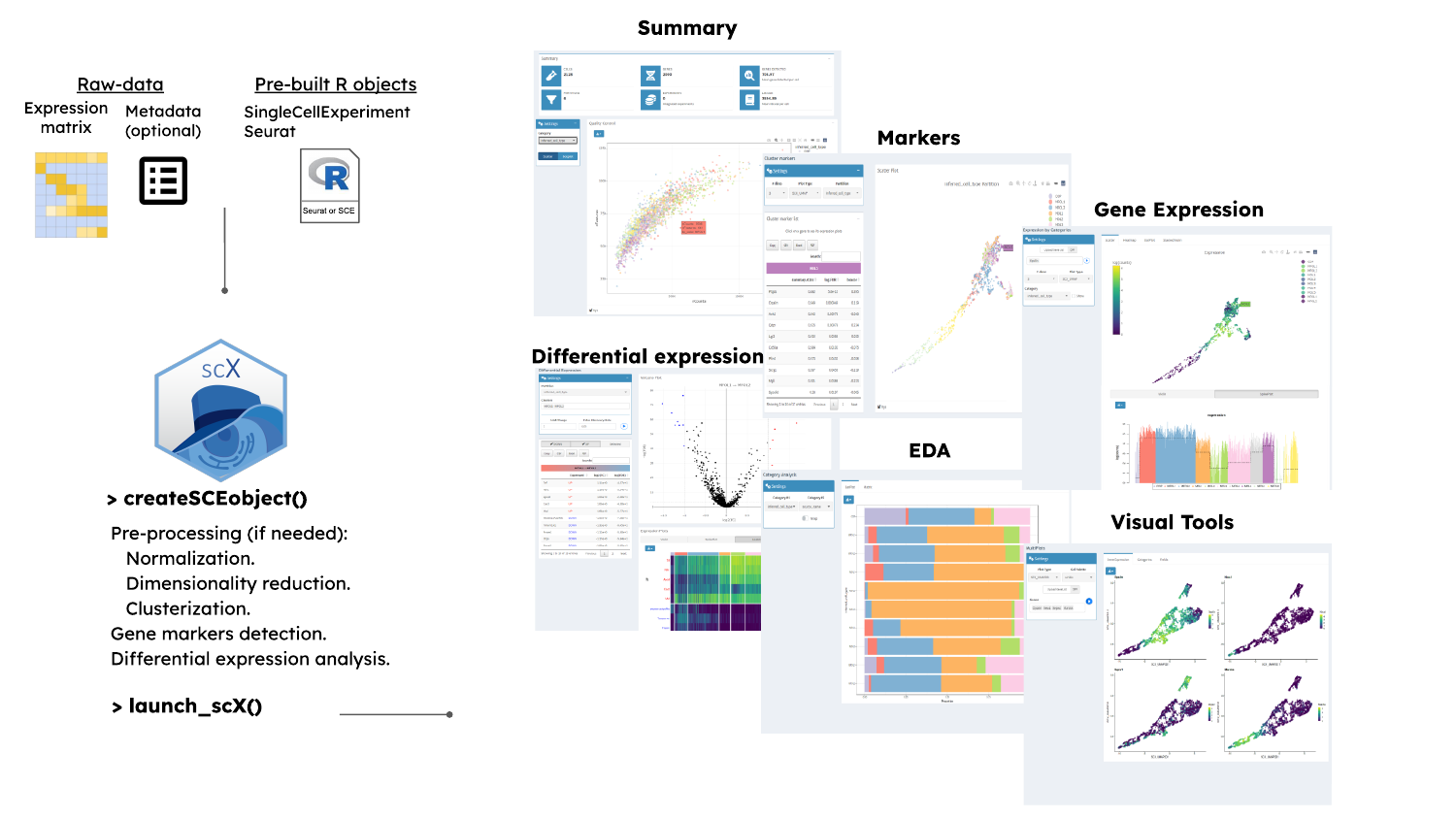}
    \end{center}
    \caption{A schematic illustration of the scX workflow is depicted on the left-hand side. The right-hand side exhibits several instances of scX's analysis and visualization capabilities.}
    \label{fig}
\end{figure*}

\section{Methods}

The scX app can be easily launched by executing two R functions (see Fig~\ref{fig}). Starting from a provided count matrix, a SingleCellExperiment object, or a Seurat object, the function 'createSCEobject' handles the offline pre-processing of the data. This function automatically executes a series of computational steps leveraging the functionality implemented in the 'scran' Bioconductor package~\cite{Lun2016scran}. 
 We adopted a normalization by deconvolution scheme~\cite{Lun2016} to eliminate systematic differences between libraries 
The identification of the most variable features involves fitting a trend on the variance vs. mean of log-normalized expression profiles 
and identifying genes with a positive biological variance component. If requested, a graph-based cell clustering procedure can be implemented. 
By default, it performs a Louvain clustering over a mutual-K nearest neighbor graph (k=20) estimated considering Euclidean cell-cell distances in the sub-space spanned by the 20 largest PCA components. It is also possible to specify any other graph-based clustering method available from the igraph package. Advanced users can directly specify an NNGraphParam object (from the bluster Bioconductor package) to achieve maximum control in the specification of this graph-based clustering task. Differential expression analysis and marker identification are conducted on one or more user-specified partitions, relying on the functionality implemented in the 'findMarkers' function. For marker identification, a Wilcoxon rank sum test is considered by default, but other options (t-test, binomial test) can also be specified. For every cluster, pairwise tests are performed against any other cluster and, by default, genes are ranked based on the maximal observed p-val (pval.type='all' in the paramFindMarkers input parameter). Optionally, the user can adopt other strategies to consolidate DE signals. Furthermore, ad-hoc pre-calculated lists of markers can also be provided.

\subsection{Summary module}
This module provides a summary of the primary descriptive details of the working dataset, such as the number of cells and genes, the mean number of genes detected per cell, etc. Additionally, it allows the visualization of the number of counts and detected features in connection with various metadata covariates, enabling the evaluation of potential batch-related issues.

\subsection{Exploratory Data Analysis module (EDA)}
This module facilitates the exploration of relationships between the covariates included in the metadata of the SCE object (specified by the 'metadataVars' and 'partitionVars' parameters of creatSCEobject). In the 'Categories' section, bar plots can be used to analyze one- or two-dimensional distribution functions involving categorical covariates. The 'Matrix' tab enables the generation of bivariate count tables. The 'Field' section of this module allows the exploration of how the value of one or more continuous covariates changes concerning another variable, which can be either numerical or categorical. Various types of plots, including box plots, heatmaps, dot plots, or stacked violin plots, can be generated to assist in this analysis.

\subsection{Markers module}
The 'Cluster markers' section facilitates the analysis of marker genes identified during the pre-processing step. When a user selects a cell displayed in the embedding window, a marker gene table is generated, including various metrics for each gene marker. By default (pval.type='all' in the paramFindMarkers input parameter of the createSCEObject function), 'summary.stats' reports the weakest observed differential expression signal between the analyzed cluster and any other cluster. 'log.FDR' is the logged largest observed FDR-corrected p-value, and 'boxcor' is the Pearson correlation value between the gene expression profile and a binary indicator vector of the cluster of interest.
The table can be saved in various formats (.csv, .xlsx, .pdf) or copied to the clipboard. Notably, clicking on a gene in the table allows the visualization of the corresponding expression field in the embedding window. Additional graphical characterizations are provided as violin and spike plots presented at the bottom of the page.

In the 'Find new markers' section, users can investigate markers for specific sets of cells, selected on the fly from 2D embeddings using the box or lasso tools (see the online manual for an animated GIF tutorial). Putative markers are identified by ranking genes in decreasing order based on their estimated boxcor values (above a minimum value of 0.3). The marker table, along with the corresponding cell list, can be downloaded. Similarly to the previous section, clicking on a marker row generates a visualization of the marker's expression pattern in the embedding dataset, and additional graphical characterizations in the form of violin and spike plots are also produced.

\subsection{Gene Expression module}
The 'Gene Expression' module facilitates the exploration of expression patterns for one or more genes of interest. Expression changes in response to different categorical and/or numerical covariates can be assessed, and coexpression patterns between pairs of genes can be analyzed.
In the 'Categories' section, one or more genes of interest can be selected (or uploaded from a file). The module displays the average expression of these genes across the embedded dataset in the 'Scatter' window. Visualization options, including heatmaps, dot plots, and stacked violin plots, are also available for analyzing the expression of these genes concerning different categorical covariates found in the metadata.
The 'Field' section allows for the analysis of gene expression in conjunction with numerical covariates that may be present in the metadata within the SCE object. This can include variables like the number of counts or pseudotime values.
Finally, the 'Co-expression' section allows for the examination of coexpression patterns between selected gene pairs in the embedding space window. The percentage of co-detection events within categorized groups of cells can also be assessed and visualized.

\subsection{Differential Expression module}
In this section, the results of the differential expression analysis can be assessed. An interactive selection of threshold values for both logFC (logarithm of fold change) and the FDR (False Discovery Rate) significance level is available. The list of differentially expressed genes is downloadable in various formats (csv, pdf, and xlsx). This section also generates a Volcano plot graphical representation, along with visualizations (violin plots, spike plots, heatmaps, and dot plots) that facilitate a more comprehensive understanding of expression patterns for up- and down-regulated genes.

\subsection{Visual tools module}
This module provides many tools to produce pdf plots with more complex or specific layouts involving gene expression patterns and covariate variables.

\subsection{Case study}
\begin{figure*}[t]
    \begin{center}
       \includegraphics[width=7.2in]{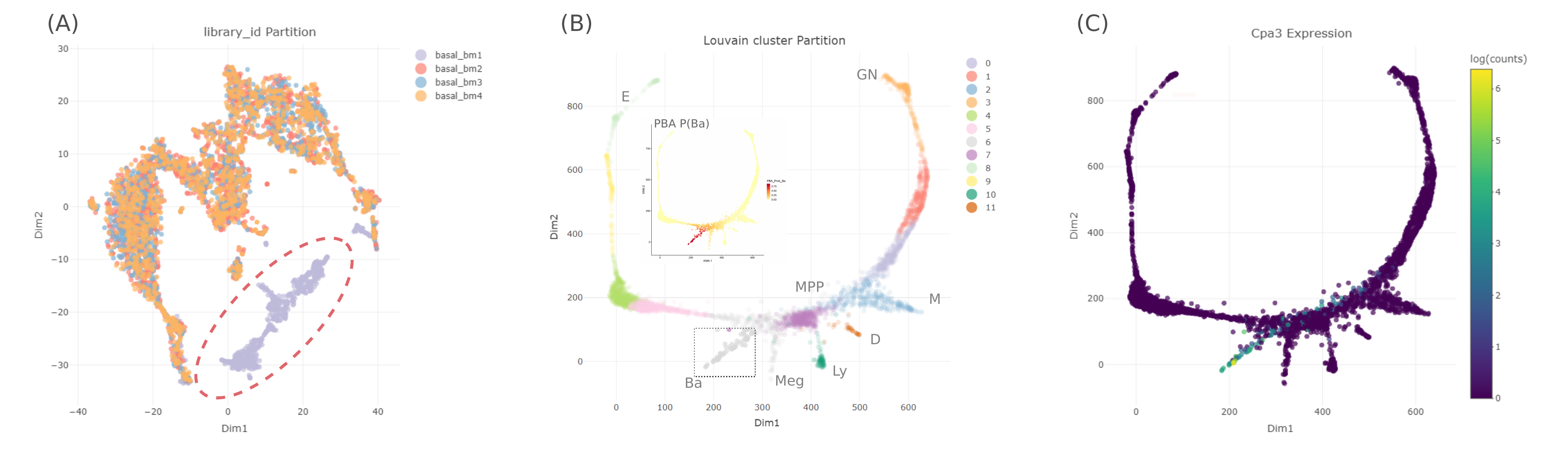}
    \end{center}
    \caption{Panel (A): 2D tSNE embedding used to highlight the large batch effect affecting 'basal\_bm1' cells. Panel (B): Force\-directed layout visualization of Tusi dataset. Louvain clusters and labels for the seven PBA inferred terminal states can be appreciated (Ba, basophilic or mast cell; D, dendritic; E, erythroid; GN, granulocytic neutrophil; Ly, lymphocytic; M, monocytic; Meg, megakaryocytic; MPP, multipotential progenitors). The dotted square schematizes the interactive cell selection process. The field of commitment probability values to the Ba state is displayed in the inset. Panel (C): Expression field of the top ranked Cpa3 putative marker gene.}
    \label{figCaseStudy}
\end{figure*}
We focused on the study conducted by Tusi and collaborators on hematopoietic lineages in mouse basal bone marrow cells~\cite{Tusi2018}. The authors performed a population balance analysis (PBA) to predict cell fate probabilities and identified seven putative commitment probabilities for each haematopoietic progenitor.
We employed scX to preprocess and re-visit their scRNAseq data (metadata and raw counts were downloaded from the paper's supporting webpage, and the used R script was included as Supplementary File). 
For our analysis, we retained the original 2D data projection (generated using a force-directed graph layout algorithm on a knn graph), and a Louvain partition of the filtered 4763 cells. Upon initial exploration a pronounced batch effect associated with 'basal\_bm1' cells, originating from a library processed in a specific sequencing run (seq\_run\_1), was identified (see Fig.~\ref{figCaseStudy}-a). Consequently, we filtered out this run, retained 4016 cells, and re-created the SCEObject for further analysis. 

We then considered the 2D force-directed layout (FDL) representation shown in Fig.~\ref{figCaseStudy}-b, where different colors were used for the 12 clusters of the original Louvain partition.  We focused on characterizing the terminal group identified as basophilic or mast cells (Ba in Fig~\ref{figCaseStudy}-b). This group exhibited high commitment probabilities to the Ba attracting state, as visualized in the $P(Ba)$ field over the graph (inset of Fig~\ref{figCaseStudy}-b). Notably, these cells were included in a broader Louvain cluster (cluster 6, gray dots in~\ref{figCaseStudy}-b). The 'find-new-marker' functionality was then employed to identify specific gene markers for this set of cells.
The top five ranked marker genes found by our tool were: Cpa3, Ms4a2, Gzmb, Cyp11a1, and Fcer1a, with boxcor values of 0.553, 0.509, 0.506, 0.496, and 0.486 respectively. The expression field of Cpa3 is shown in Fig.~\ref{figCaseStudy}C. 
To validate these putative markers we referred to the work of Miao and collaborators, who developed the SCCAF strategy (Single-Cell Clustering Assessment Framework) for cell type discovery from single-cell expression data~\cite{Miao2020}. We found that the first three marker genes identified by scX were also top-ranked features in SCCAF's logistic regression model for recognizing the Ba cell group (cluster 10 in Fig5-d of~\cite{Miao2020}). Additionally, we found further support for the complete set of scX-identified markers in several bibliographic references~\cite{SiddhurajOA4848, Metcalfe2016, Miyake2023, Silva-Gomes2021, Hiroyasu2021}. 

\subsection{Computational Demands}
To test the computational demands of a typical scX pipeline we considered the mouse nervous system's scRNA-seq data from Zeisel et al.(2018). Running-time and peak memory consumption tests were conducted on an Intel(R) Xeon(R) Silver 4116 CPU @ 2.10GHz, 514G RAM server, considering incrementally subsampled datasets. Details and results are summarized in Table ST2 (included as supplementary material). The most demanding step was data preprocessing (createSCEobject function), requiring a peak of 24Gb of RAM and 180 minutes for N=160,000 cells. We found that peak memory usage and running time scaled linearly at a rate of 1.3Gb/10k-cells and 10.4min/10k-cells, respectively (see Figure SF1 in Sup.Mat.). It's important to note that the number of partitions and clusters can influence the processing time of a given dataset. For visualization purposes a subsampling strategy can be specified at pre-processing time (default setting of 50,000 cells) to ensure a smooth interactive experience.

\section{Conclusions}
We developed scX, a Shiny-based application that enhances collaboration between bioinformaticians and experimental biologists in joint projects. The platform is user-friendly and highly interactive, fostering a collaborative environment that could significantly advance the development of joint projects. 


\section{Code availablity}
Source code can be downloaded from \url{https://github.com/chernolabs/scX}. User manual available at \url{https://chernolabs.github.io/scX/}. A docker image is available from dockerhub as chernolabs/scx.

\section*{Acknowledgements}
D.G., A.C., and A.F.S. are investigators in the Consejo Nacional de Investigaciones Científicas y Técnicas (CONICET). N.B.R, A.A.B. M.H. and M.B. were supported by CONICET fellowships. This work was supported by grants the National Institute of Neurological Disorders and Stroke (NINDS) and Fogarty International Center (FIC) (R01NS103758) to P.A. and A.F.S., and the Argentine Agency for the Promotion of Science and Technology (PICT-2020-0046 and PICT-2021-0077) to A.F.S., (PICT 2018-03713) to A.C. and M.S.B. (postdoctoral fellowship) and (PICT 2017-0389) to D.G.




\end{document}